\documentclass[12pt]{article}
\usepackage{graphicx}
\usepackage{amsmath,amssymb,amsfonts}
\usepackage{mathrsfs}
\usepackage{latexsym}
\usepackage{bm}
\usepackage{bbm}
\usepackage{cite}
\usepackage{subfigure}
\usepackage{mathrsfs}
\usepackage{indentfirst}
\usepackage{amsmath}
\usepackage{amssymb}
\usepackage{cases}

\usepackage[colorlinks]{hyperref}
\hypersetup{citecolor=blue,linkcolor=blue,urlcolor=blue}

\begin{document}

\title{\Large\bf
Involution game with specialization strategy}

\vspace{0.3truecm}
\author{
 Bo Li\footnote{Email: libo312@mails.ucas.ac.cn}
  \\ \\
 \textit{PKU-WUHAN Institute for Artificial Intelligence},\\
  \textit{Wuhan, 430070, P.~R.~China}
}
\date{\today}

\maketitle
\begin{abstract}
Involution now refers to the phenomenon that competitors in the same field make more efforts to struggle for limited resources but get lower individual ``profit effort ratio''.
In this work, we investigate the evolution of the involution game when competitors in the same field can adopt not only the strategy of making more efforts but also a specialization strategy which allows competitors to devote all their efforts to part of the competitive field. Based on the existing model, we construct the involution game with the specialization strategy and simulate the evolution of it on a square lattice under different social resource, allocation parameter (characterizing the intensity of social competition), effort and other conditions. In addition, we also conduct an theoretical analysis to further understand the underlying mechanism of our model and to avoid illusive results caused by the model settings. Our main results show that, when the total effort of the specialization strategy and the ordinary strategy is equal, the group composed of all the agents has a certain probability to choose the ordinary strategy if the allocation parameter is very large (that is to say the intensity of competition is very weak), otherwise the group will choose the specialization strategy; when the total effort of the two strategies is not equal, the proportion of the specialization strategy adoption is related to the social resource, the effort and the allocation parameter. To some extent, our study can explain why division of labor appears in human society, and provide suggestions for individuals on competition strategy selection and governments on competition policy development.
\end{abstract}

\newpage

\section{Introduction}\label{sec_intro}
The term ``involution'' was first proposed by Kant, a German philosopher, in his book \textit{Critique of Judgment} \cite{kant1987critique,wang2021replicator}.
The concept of involution is developed by two anthropologists, one of whom is Alexander Lovich Goldenhitzer, he called the phenomenon that a kind of cultural mode has reached a certain final form, can neither stabilize nor change into a new form, but can only become more complex internally as ``involution'' \cite{mccullough2019review}.
The other anthropologist is Geertz and he used the word ``involution'' in his book \textit{Agricultural Involution: The Processes of Ecological Change in Indonesia} \cite{geertz1963agricultural,wang2021replicator}, which tells about what he found in Indonesia. 
Geertz discovered that the agriculture in island Java could not be expanded outward due to the lack of capital, limited land quantity and administrative barriers, ultimately leading to the increasing labor force constantly filling the limited rice production. In this situation the internal structure of agriculture becoming more sophisticated and complex, thus forming ``growth without development''. Geertz summarized this process with ``agricultural involution''.
In modern society, the competitions in many fields are becoming increasingly fierce and everyone invests more and more effort, but these fields show the characteristics of involution—growth without development.
After the word ``involution'' becomes popular on the Internet, it emphasizes the ineffective competition and the decline of income-effort ratio.

Faced with involution, besides adopting the strategy of making more effort that will exacerbate the involution, individuals and organizations can also use specialization strategies to obtain more resources and income \cite{romer1987growth,williams2018specialization,krugman1996integration}. Generally, specialization strategies refer to concentrating all efforts and capabilities in a narrow area, for companies the specialization strategy may means focusing on the production of a very limited range of products or services; for restaurants it may means making a certain type of food like KFC or Pizza Hut; and for athletes it means engaging in only a single sport, and so on. 
Specialization strategy or differentiation strategy is common in business competitions, there are some scholars who have done some researches on them, such as
Norton et al. \cite{norton1993specialization}, Sorenson \cite{sorensen2008learning}, Gompers et al. \cite{gompers2009specialization} and Hochberg et al. \cite{hochberg2015specialization} studied the problems related to the specialization strategy in the venture capital industry;
Boehe and Barin \cite{boehe2010corporate}, Banker et al. \cite{banker2014does}, and Semuel et al. \cite{semuel2017effect} studied the relationship between the differentiation strategy and company performance.

The intensification of social competition is an important driving force for competitors to adopt the specialization strategies, which leads to the increasingly sophisticated division of labor \cite{chen1987origin,norton1993specialization,yang1998specialization,ricoy2003marx,metcalfe2003equilibrium,ruiz2004markets,siebert2006locational,sorensen2008learning,gompers2009specialization,hochberg2015specialization,barker2016asymmetry}. 
Most of the related researches are based on economic and social perspectives, which belong to the research scope of economics or sociology, such as Ref. \cite{metcalfe2003equilibrium} and Ref. \cite{siebert2006locational}. The topic of Ref. \cite{metcalfe2003equilibrium} is competition and policy on science and technology, and the Ref. \cite{siebert2006locational} studies the locational competition in the international division of labor.
There are also some researches based on theoretical modeling, for example, Chen \cite{chen1987origin} established a mathematical model to try to explain the origin of division of labor, and concluded that random mechanism plays a very important role in the process of social evolution.

Against the backdrop of widespread attention to the phenomenon of involution, Wang et al. \cite{wang2021replicator,wang2022modeling} proposed an involution game model on a square lattice, in which agents can choose between two strategies with making different effort to strive for resources. 
The agents in the model adjust their strategies according to the payoff difference with neighbors, and the ratio of adopting the strategy of making more efforts is used as the indicator to measure the degree of involution. 
Based on the above settings, they found that more abundant social resources lead to the involution; an increase in allocation parameter (decrease in the intensity of competition) in resource distribution suppresses the involution; and an increase in the cost of more effort does not always aggravate or suppress the involution. 
The conclusions may be helpful to resist involution. 
In another work, Wang et al. \cite{wang2022involution} also discussed the involution model when social resources have spatiotemporal heterogeneity, and  in this work they changed the resource allocation mechanism.
In other academic communities such as psychology \cite{liu2022have}, there has also been some researches on involution not long ago.

One of the methods involved in the works of Wang et al. is the spatial evolution game \cite{nowak1992evolutionary,wang2021replicator,wang2022modeling,wang2022involution,perc2013evolutionary,perc2017statistical,yang2016critical,tian2023influence}, which was first proposed by Nowak and May and has been widely used \cite{nowak1992evolutionary}. 
Another main method is Agent-based model, which has been widely used in financial market simulation \cite{lux1999scaling,challet2000modeling,shi2019pyramid,shi2021pyramid}, social simulation \cite{li2008agent,bianchi2015agent} and various other simulation studies \cite{heath2009survey,bruch2015agent}. 

Inspired by the specialization strategy in real world and the works on involution game of Wang et al. \cite{wang2022modeling}, we investigate the evolution of the involution game that agents can take not only the strategy of making more effort but also a specialization strategy which allows agents to devote all their effort to part of the resources.
Compared with the existed involution model, we redefine the involution game, creating a new pair of strategies akin to the classical one, which offers valuable insights into involution from a different angle.
Consequently, the central question shifts toward ``whether to choose the specialization strategy'' and ``the emergence of labor division'' rather than the classic model's focuses on ``how to decrease involution''.
We simulate the evolution of the involution game on a square lattice under different social resource, allocation parameter, effort and other conditions.
To further understand the underlying mechanism of our model and avoid illusive results caused by the model settings, we also conduct a theoretical analysis, which is consistent with the simulation results well.

The innovation and contribution of this study are that, we construct a new involution model with specialization strategy, which enriches the relevant researches about involution, homogeneous competition and strategy selection.
Although this work is based on the existing involution game model, it employs a novel setting to explain why human social division of labor is becoming increasingly refined, and give suggestions on the social conditions under which the specialization strategies are appropriate.
For applications, our model show that, in industries with low level of competition intensity, it is suitable to adopt the ordinary strategy with low investment when the resource is poor, while the resources are abundant the choice of strategy is best according to the specific situation of the degree of competition and the total amount of the resources;
in industries with high level of competition intensity, our model show that, unless the total amount of social resources is very low, we should adopt the specialization strategy with relatively small investment, which can achieve more resources and make less effort.

The remainders of the paper are organized as follows: 
In Sec.~\ref{sec_model} an involution evolution model with specialization strategy is proposed;
Sec.~\ref{sec_result} presents the simulation results and related analysis;
Sec.~\ref{sec_ana} conducts an theoretical analysis to avoid illusive results and mine the underlying mechanism;
some discussion and conclusion are given in Sec.~\ref{sec_con}.

\section{Model}\label{sec_model}

Compared to ordinary strategies, the essential feature of specialization strategies is to put all efforts into a relatively small field. Now let us consider the simplest and most representative case, that is, there are two subdivisions, and use a two-dimensional vector $(E_1, E_2)$ to represent the effort of a competitor investing in two fields.
Without losing representativeness, set $E_1=E_2$ corresponds to the ordinary strategy, which means the competitor invests the same effort in both fields; 
set one of $E_1$ or $E_2$ equaling to 0 corresponds to the specialization strategy, which indicates the competitor only invests effort in one field.

To study the performance and evolution of these two strategies in the involution dilemma, we consider an involution game on an $L\times L$ square lattice with periodic boundary conditions which is actually a regular network with degree $D=4$, and there are $N=L^2$ agents are distributed on the nodes. 
Different from the settings in the existed models \cite{wang2022modeling,wang2022involution}, we assume agents can take the specialization strategy described above, which allows each agent devotes all his efforts into striving for two parts of the whole resource $M$, $M_1$ and $M_2$. 
Considering the heterogeneity of agents and changes in time,
we denote the strategy $S$ of agent $i$ at time step $t$ as a vector $(E_1(i,t), E_2(i,t))$, where $E_1(i,t)$ and $E_2(i,t)$ represent the effort invested in $M_1$ and $M_2$ respectively.
As done in Ref.~\cite{wang2022modeling}, we also adopt the Boltzmann distribution in statistic physics for the social resource allocation.
This issue has been sufficiently discussed in Ref.~\cite{wang2022modeling}, and the Boltzmann distribution means the entropy $E=-\sum_{j\in \mathbb{N}(k)} P(j,t) \ln{P(j,t)}$ takes its maximum, where $P(j,t)$ is the proportion of the social resource $M$ in the social group $\mathbb{N}(k)$ allocated to an agent $j$ taking effort $\left(E_1(j,t), E_2(j,t)\right)$.
We follow this practice then the payoff $\pi (i,t)$ of agent $i$ at time step $t$ is given by
\begin{equation}
	\pi (i,t) = \sum_{k\in \mathbb{N}(i)} \left\{ \frac{e^{\frac{E_{\scriptscriptstyle 1 }\left(i,t\right)}{\kappa_1}}}{\sum\limits_{j\in 
			\mathbb{N} \left( k \right)}e^{\frac{E_1\left(j,t\right)}{\kappa_1}}}\cdot M_1+
	\frac{e^{\frac{E_{\scriptscriptstyle 2 }\left(i,t\right)}{\kappa_1}}}{\sum\limits_{j\in 
			\mathbb{N} \left( k \right)}e^{\frac{E_2\left(j,t\right)}{\kappa_1}}} \cdot M_2 - \left [ E_1\left(i,t\right)+E_2\left(i,t\right) \right ] \right\},
\end{equation}
where $\mathbb{N}(i)$ is the node collection including node $i$ and its neighbors, and the meaning of $\mathbb{N}(k)$ is similar.
The parameter $\kappa_1$ refers to the allocation parameter: the smaller the value, the more dependent the resource allocation is on the effort made by the agent; the larger the value, the amount of effort has little effect on the resources allocation, on the contrary more efforts will bring higher costs and lead to lower payoff \cite{wang2022modeling}.
For $M_1$ and $M_2$, in order to simplify without loss of generality, we will set $M_1=M_2=M/2$ in the simulations.

The proportion of the specialization strategy adoption is the most concerned indicator during the evolution of the involution model. 
We first study the indicator in the involution game on the assumption that the specialization strategy and the ordinary strategy compete for the resources with the same effort. 
To this end, we assume that the strategy of agent $i$ can only be chosen from the strategy $(E/2, E/2)$ and the strategy $(E, 0)$ or $(0, E)$, where the strategy $(E/2, E/2)$ is considered as an ordinary strategy, and the strategy $(E, 0)$ or $(0, E)$ is considered as a specialization strategy. 
In addition, if agent $i$ chooses to adopt the specialization strategy due to the influence of neighboring agents, he will randomly choose between $(E, 0)$ and $(0, E)$ with equal probability.
It is worth noting that in our setting, this randomness only exists when the agent transitions from the ordinary strategy to the specialization strategy.
Assuming that a certain agent's strategy at time step $t$ is $(E, 0)$, its strategy will not randomly change to $(0, E)$ at time step $t+1$, and the agent can only transform the strategy into the ordinary strategy $(E/2, E/2)$ due to the influence of neighboring agents.

In the simulations, we also relax the assumption that the total effort of all the agents is the same, and represent the ordinary strategy as $(\alpha E/2, \alpha E/2)$, where $\alpha > 0$. By this way, the effort ratio of the ordinary strategy to the specialization strategy is $\alpha$.

Up to now, the agents in above involution game still lacks a strategy adjustment mechanism. Here we follow a regular approach from spatial evolutionary game theory, just like that in Ref.~\cite{szabo1998evolutionary,blume1993statistical,vukov2006cooperation,traulsen2007pairwise,szolnoki2009selection,traulsen2010human,wang2022modeling,wang2022involution}. 
Specifically, for an agent $i$ in the first step, it randomly selects an agent $j$ from its neighbors, where $j \in \mathbb N (i)\setminus i$. 
Next, agent $i$ takes the strategy of agent $j$ at time step $t$ as its own strategy at time step $t + 1$ with a probability formulated by the Fermi–Dirac function
\begin{equation}
	p\left[ S(i,t+1) \gets S(j,t) \right] = \frac{1}{1+e^{\frac{\pi(i,t)-\pi(j,t)}{\kappa_2}}}, \label{eq_adj}
\end{equation}
where the parameter $\kappa_2$ describes probability to the payoff difference $\pi(i,t)-\pi(j,t)$. 
$\kappa_2$ is critical in updating strategies, the lower the value, the more agent $i$ tends to adopt $j$'s strategy which bring a higher payoff to $j$, if $\kappa_2$ is lower the opposite is true.
When $\kappa_2$ is too large, payoff plays no role in strategy updating.
In this work, we take $\kappa_2 = 0.1$ as in the earlier work \cite{wang2022modeling}.

\section{Results and discussion}\label{sec_result}
In the simulation, the square lattice is set to $50\times50$ so there are 2500 agents in total and each agent has 4 neighbors. 
At the beginning, each agent is randomly assigned the specialization strategy and the ordinary strategy with equal probability (50\%). 
For agents whose initial strategy is the specialization strategy, the probability of choosing either $(E, 0)$ or $(0, E)$ is 50\% each.
Using synchronous or asynchronous update rules will not have an essential impact on the simulation results, for convenience we use synchronous update rules in this work.

The focus variable of the study, denoted by $F_S$, is the fraction of the agents taking the specialization strategy in $L\times L$ agents, which measures the tendency of the group to select the specialization strategy.
In the following, we first study the situation of all agents making equal effort, and then study the situation of agents with different strategy making unequal effort. 
The first situation is to explore which strategy is more advantageous when the effort of the specialization and the ordinary strategy is the same, while the second situation is to explore which strategy is more advantageous when the effort of the specialization and the ordinary strategy is different. 
For these two cases, we use different methods to determine the evolution results of $F_S$.

\subsection{All agents making equal effort}
\begin{figure}[!htbp]	
	\centering
	\includegraphics[scale=0.25]{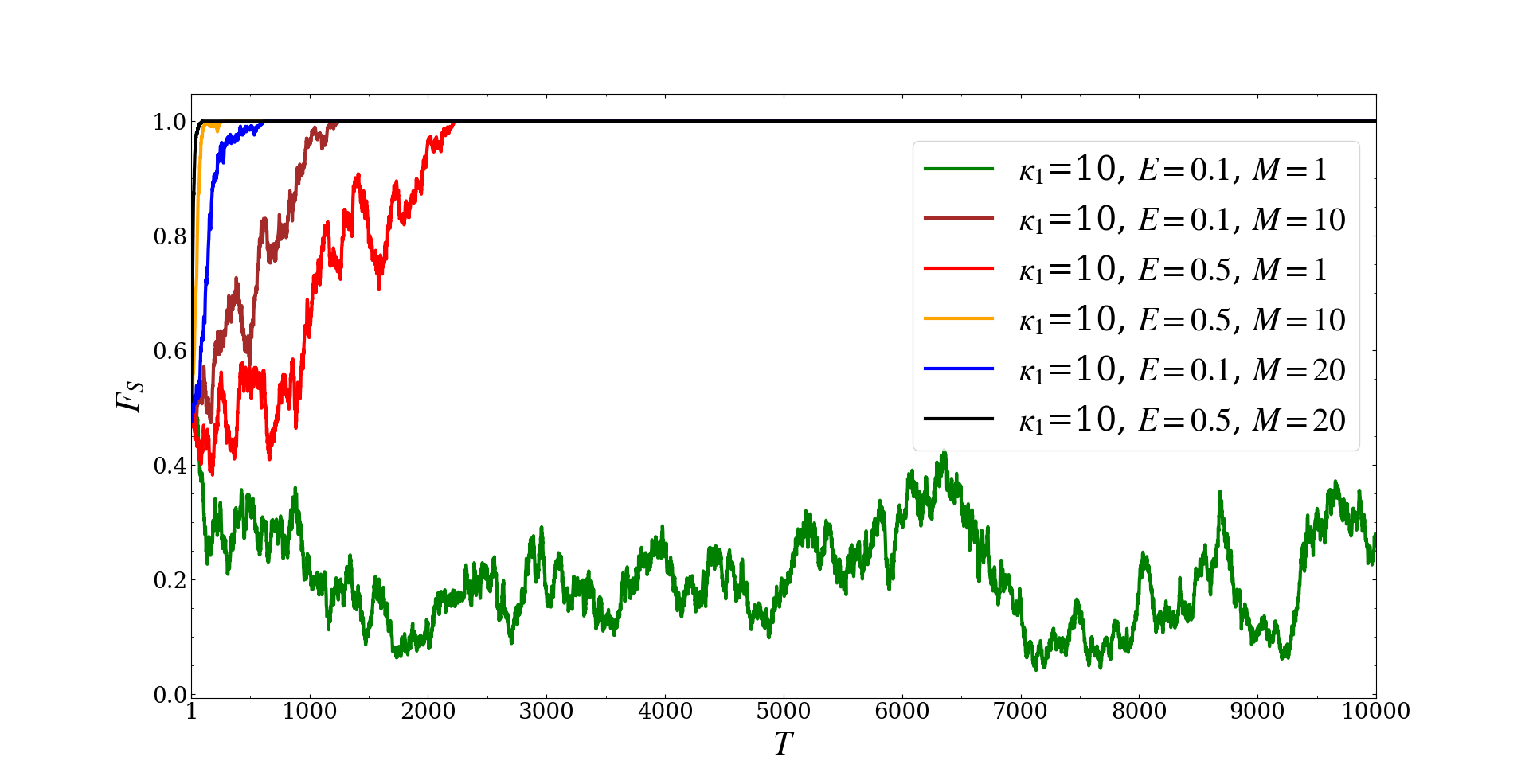}
	\caption{Evolution of $F_S$ over time step $T$ under different parameter combinations when all agents making equal effort, which are $\kappa_1=10$, $E=0.1$, $M=1$; $\kappa_1=10$, $E=0.1$, $M=10$; $\kappa_1=10$, $E=0.5$, $M=1$; $\kappa_1=10$, $E=0.5$, $M=10$; $\kappa_1=10$, $E=0.1$, $M=20$; and $\kappa_1=10$, $E=0.5$, $M=20$.}
	\label{fig_t}
\end{figure}

First of all, we find that the evolution results of $F_S$ will emerge random phenomena under some parameter conditions, while under other parameter conditions the results are all the agents adopt the specialization strategy. 
Randomness here means that $F_S$ fluctuates between 0 and 1, and does not reach non-stationary equilibrium as the green line shown in Fig.~\ref{fig_t}.
From the figure, we can find that random phenomena occur when both $E$ and $M$ are relatively small if $\kappa_1$ is fixed.
It must be pointed out that the reason why $F_S$ reaches 0 or 1 is not that the selected lattice number is too small, but that the specialization strategy has absolute advantage in the evolution process.
In fact, we once took some larger $L$, but the results are not fundamentally different.

The occurrence of random phenomena can be understood as follows: when $E$ is approaching 0, the difference between the ordinary strategy and the specialization strategy is very small, so in the simulation results, $F_S$ fluctuates randomly between 0 and 1.
In subsection~4.1, we will further conduct a theoretical analysis of it.
Because of the appearance of randomness, we repeat the same parameter combination for 20 times then take the average of $F_S$ at the time step $T=3000$.

\begin{figure}[!htbp]	
	\centering
	\includegraphics[scale=0.25]{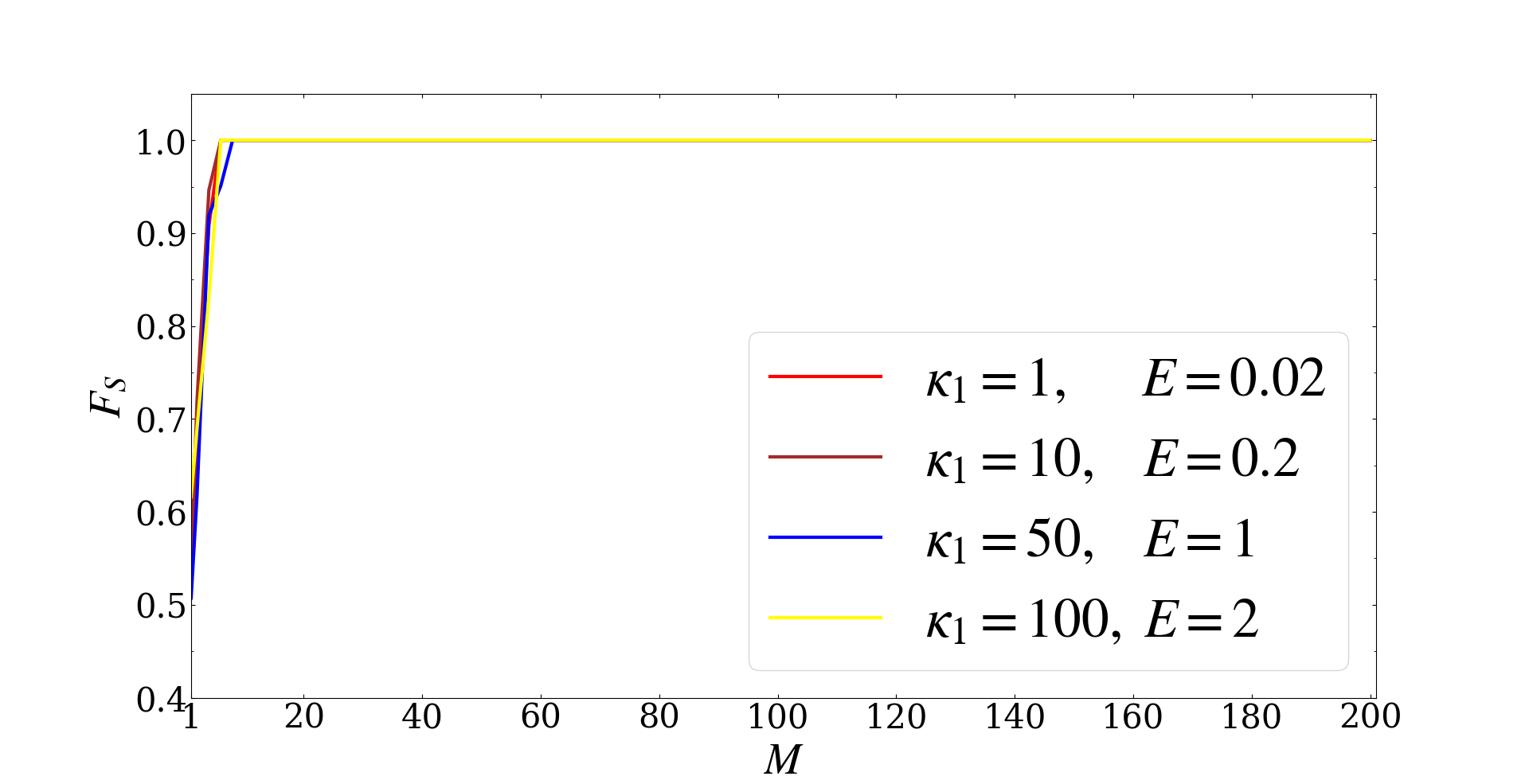}
	\caption{The relationship of $F_S$ with $M$, where the other parameters are: $\kappa_1=1, E=0.02$; $\kappa_1=10, E=0.2$; $\kappa_1=50, E=1$ and $\kappa_1=100, E=2$. Each parameter combination has been repeated 20 times, and $F_S$ is the average of all the simulation results.}
	\label{fig_phen}
\end{figure}

\begin{figure*}
	\centering
	\subfigure[]{\label{fig_k1}
		\includegraphics[width=0.43\linewidth]{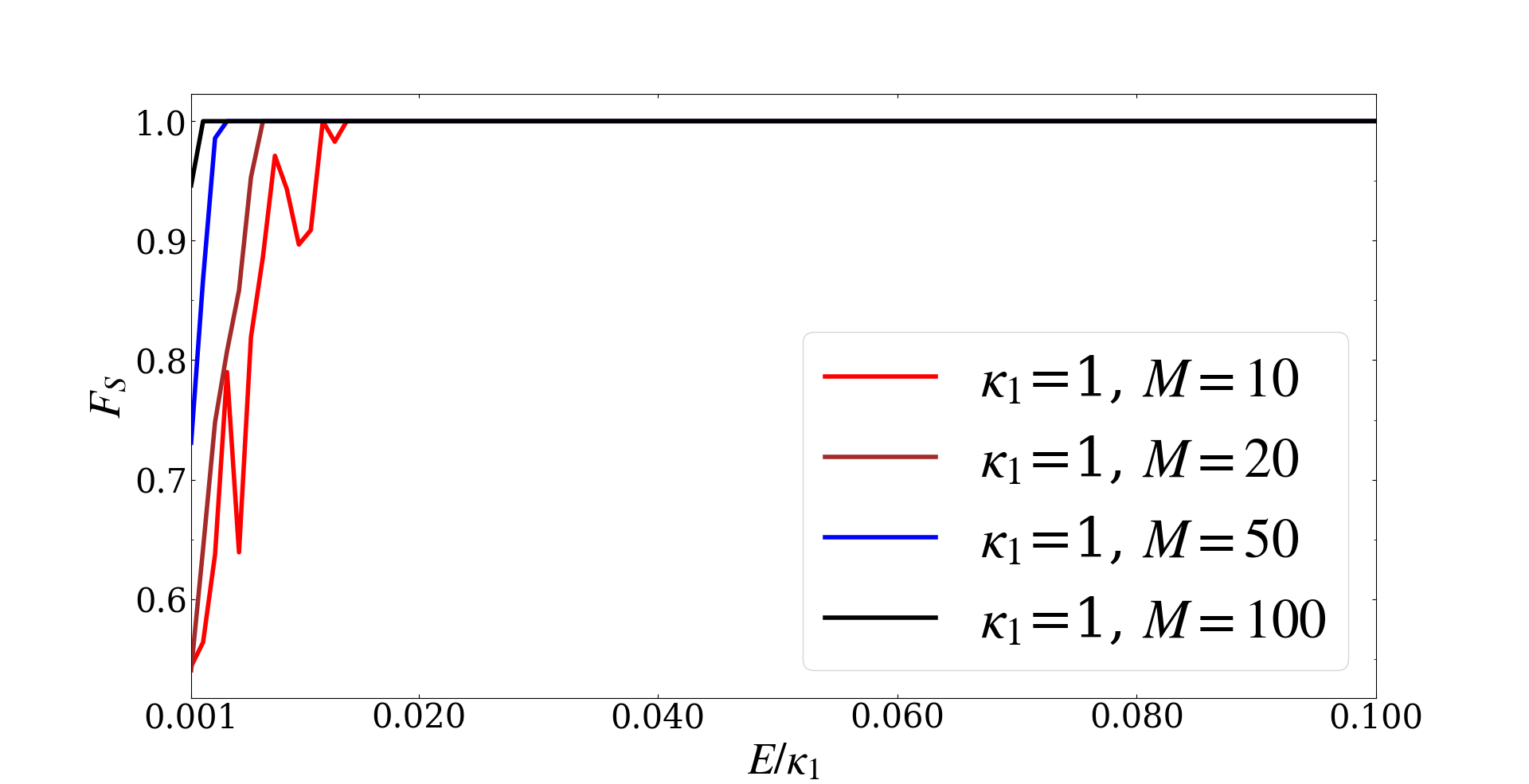}}
	\hspace{0.01\linewidth}
	\subfigure[]{\label{fig_k10}
		\includegraphics[width=0.43\linewidth]{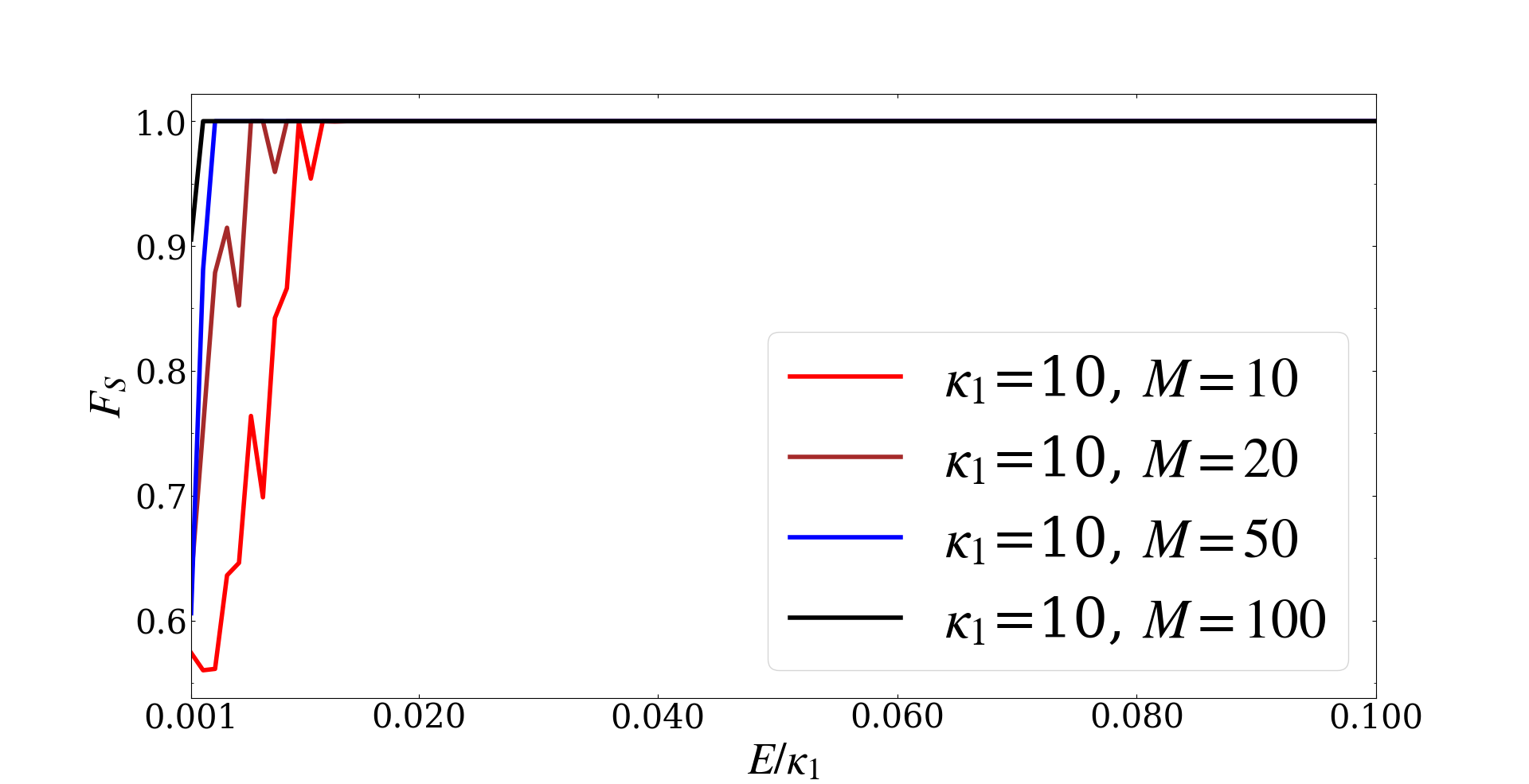}}
	\vfill
	\subfigure[]{\label{fig_k20}
		\includegraphics[width=0.43\linewidth]{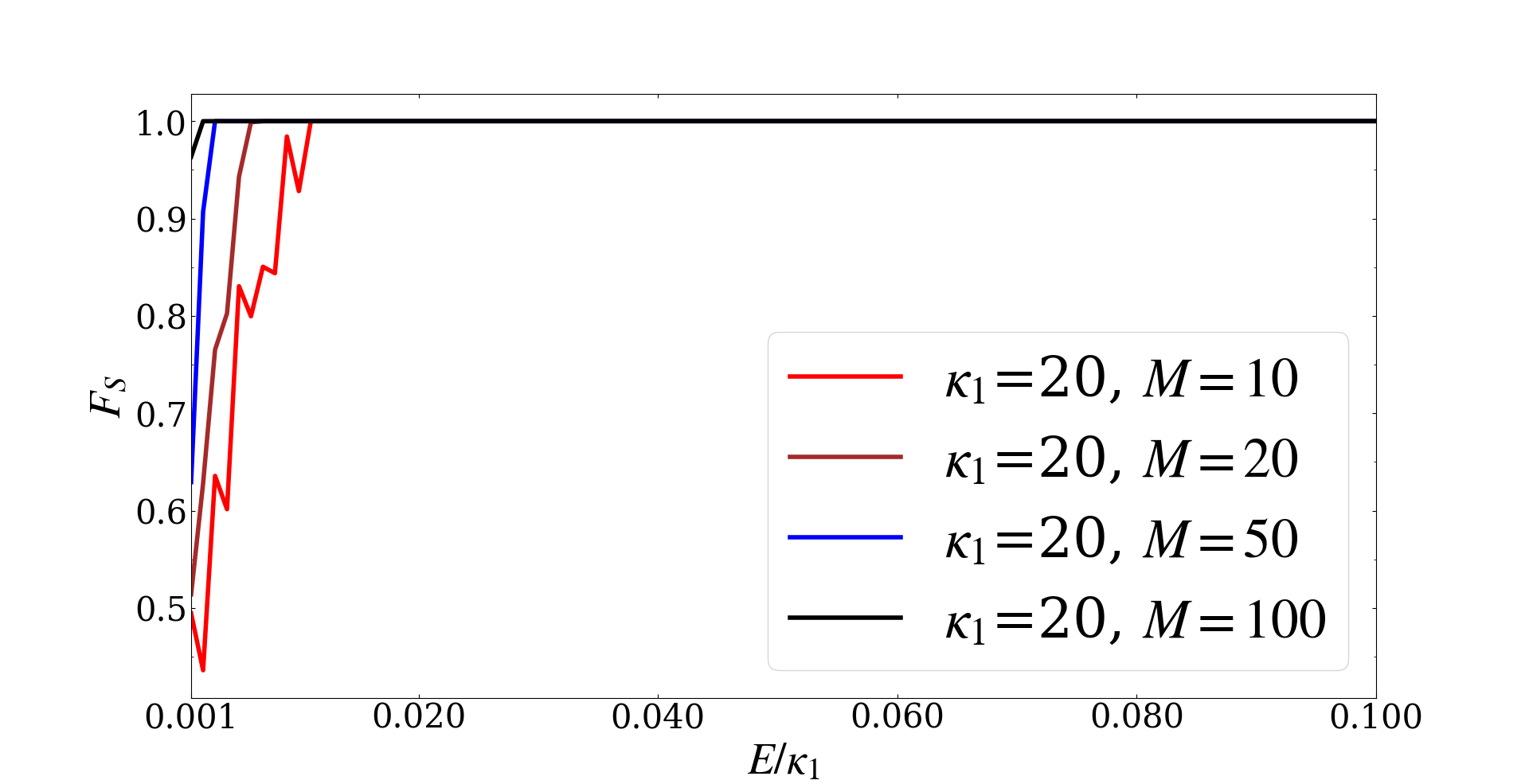}}
	\hspace{0.01\linewidth}
	\subfigure[]{\label{fig_k50}
		\includegraphics[width=0.43\linewidth]{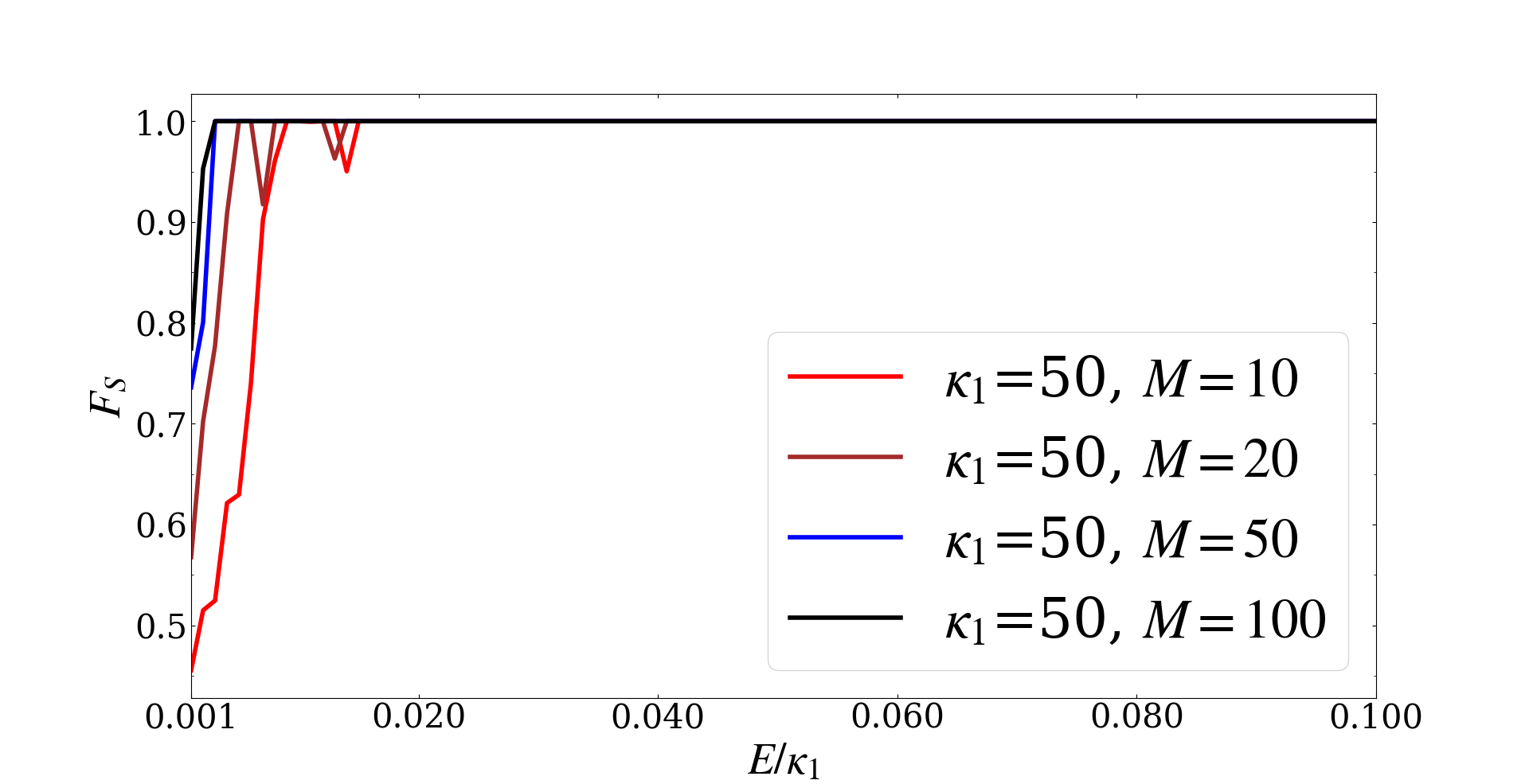}}
	\caption{The relationship of $F_S$ with $E/\kappa_1$, where the combinations of the other parameters are: (a) $\kappa_1=1$, $M=10$, $M=20$, $M=50$ or $M=100$; (b) $\kappa_1=10$, $M=10$, $M=20$, $M=50$ or $M=100$; (c) $\kappa_1=20$, $M=10$, $M=20$, $M=50$ or $M=100$; (d) $\kappa_1=50$, $M=10$, $M=20$, $M=50$ or $M=100$. Each parameter combination has been repeated 100 times and the average of $F_S$ is taken.}
	\label{fig_Ek}
\end{figure*}

Fig.~\ref{fig_phen} shows that $F_S$ will quickly close to 1 as $M$ increases when $E/\kappa_1$ is small, which indicates that the specialization strategy will be almost certainly adopted by the whole group with abundant resources under the condition that $E/\kappa_1$ is small.
To further study this phenomenon, we take $E/\kappa_1$ as a variable to simulate the evolution of $F_S$ under different parameter combinations, and the averages of $F_S$ are presented in Fig.~\ref{fig_Ek}. As can be seen from the figure, under different combinations of the allocation parameter $\kappa_1$ and the total resource $M$, a small change of $E/\kappa_1$ near 0 can cause a sharp change of $F_S$. 
That is to say, with large $E/\kappa_1$, the specialization strategy is almost certain to be adopted by the whole group than just partial members.
This phenomenon can be called phase transition \cite{szabo2002phase,szolnoki2011phase}, because with the small change of $E/\kappa_1$, the group of the agents has changed from having some probabilities to choose the ordinary strategy to almost certainly adopt the specialization strategy.
It can also be seen from Fig.~\ref{fig_Ek}, the larger the resource $M$, the more intense the transition.

The above simulation results can be explained as follows: if all the agents in competitions pay the same effort and the value of the effort is very small compared with the allocation parameter, the richer the resources, the agents are more inclined to adopt the specialization strategy, and when $M$ exceeds a certain value, the specialization strategy will be completely dominant in the population of the participants;
if all agents in competitions pay the same effort but the value of the effort is relatively large compared with the allocation parameter (such as $E/\kappa_1>0.1$), the evolution results in this case are that all the participants are almost certain to adopt the specialization strategy.
In addition, if all participants pay the same effort and the ratio of the effort $E$ to the allocation parameter $\kappa_1$ is approaching 0, the averages of $F_S$ will change dramatically with the ratio, which is a phase transition about the strategy selection of the group. 
Besides, the transition becomes more rapid with the increase of the resource $M$.

\subsection{Agents with different strategy making unequal effort}
Next, we study the situation that the effort of the specialization strategy and the ordinary strategy is different. 
We represent the ordinary strategy as a two dimensional vector $(\alpha E/2, \alpha E/2)$ and the specialization strategy as $(E, 0)$ or $(0, E)$, so the effort ratio of the ordinary strategy to the specialization strategy is $\alpha$.

\begin{figure}[!htbp]	
	\centering
	\includegraphics[scale=0.25]{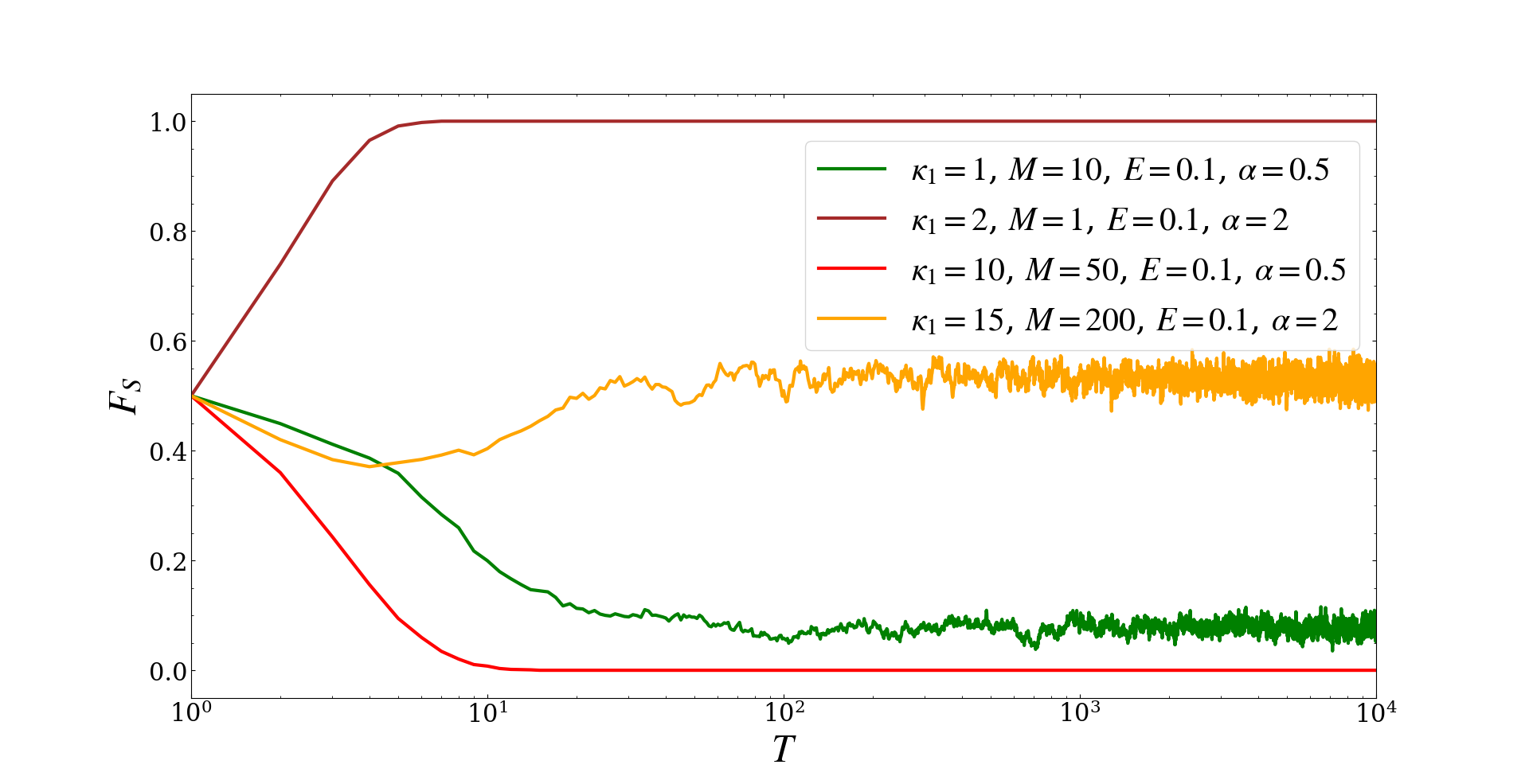}
	\caption{Evolution of $F_S$ over time step $T$ under different parameter combinations when agents with different strategy making unequal effort, the parameter combinations are $\kappa_1=1$, $M=10$, $E=0.1$, $\alpha=0.5$; $\kappa_1=2$, $M=1$, $E=0.1$, $\alpha=2$; $\kappa_1=10$, $M=50$, $E=0.1$, $\alpha=0.5$; and $\kappa_1=15$, $M=200$, $E=0.1$, $\alpha=2$.}
	\label{fig_t_2}
\end{figure}

In this situation, $F_S$ will reach an unbalanced steady state under some parameter combinations and $0<F_S<1$, while under other parameter combinations $F_S$ will reach 0 or 1 quickly, as shown in Fig.~\ref{fig_t_2}. It must be pointed out that the reason why $F_S$ reaches 0 or 1 is not that the selected lattice number is too small, but that the specialization strategy or the ordinary strategy has absolute advantage in the evolution process.
From Fig.~\ref{fig_t_2}, it can be seen that when the time step $T$ is above 3000, $F_S$ either remains at 0 or 1, or reaches non-stationary equilibrium. When in non-stationary equilibrium, we use the mean of $F_S$ from 3000 to 5000 steps as the final state of $F_S$.

\begin{figure}[!htbp]	
	\centering
	\includegraphics[scale=0.25]{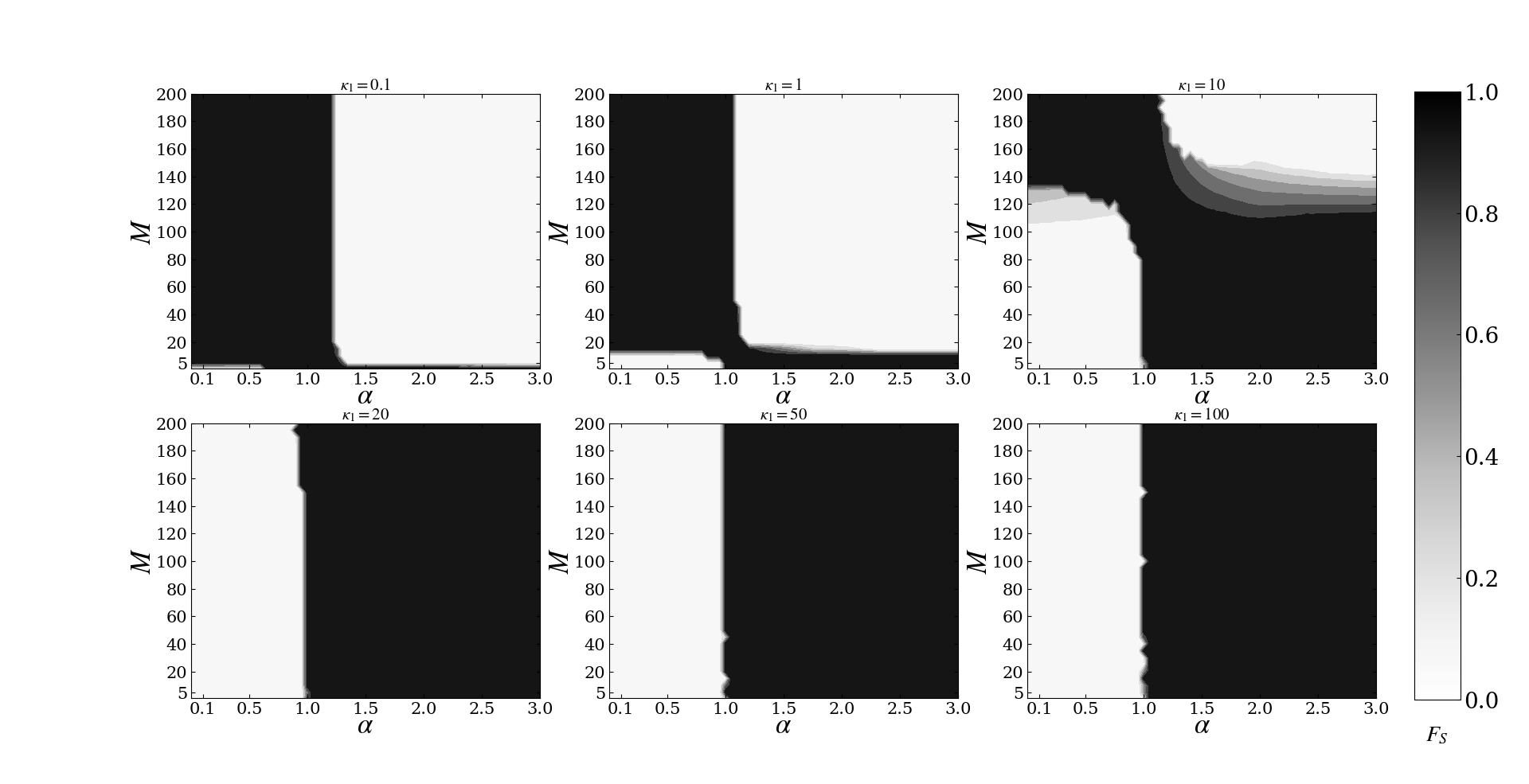}
	\caption{The proportion of the group adopting the specialization strategy, $F_S$, as a bivariate function of the resource $M$ and the ratio $\alpha$ when $E=0.1$ and $\kappa_1=0.1, 10, 20, 50, 100$.}
	\label{fig_m_a}
\end{figure}

\begin{figure}[!htbp]	
	\centering
	\includegraphics[scale=0.25]{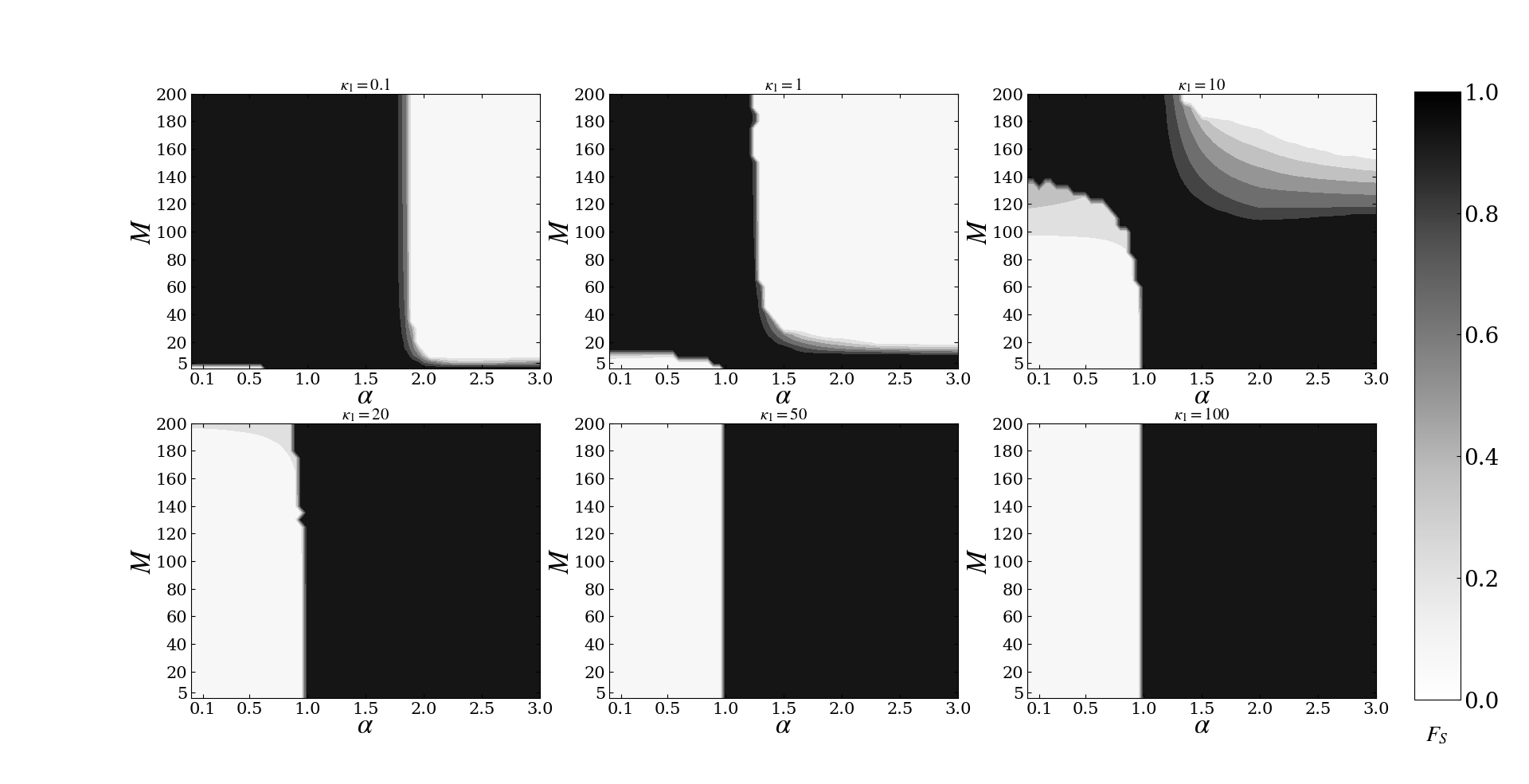}
	\caption{The proportion of the group adopting the specialization strategy, $F_S$, as a bivariate function of the resource $M$ and the ratio $\alpha$ when $E=1$ and $\kappa_1=0.1, 10, 20, 50, 100$.}
	\label{fig_m_a_e1}
\end{figure}

In the Fig.~\ref{fig_m_a}, we show the relationship of $F_S$ with the resource $M$ and the ratio $\alpha$ when the allocation parameter takes different values and $E=0.1$. 
As can be seen from the figure, with $\kappa_1$ increasing, the area on the left side of the subgraphs where the specialization strategy is dominant decreases, and the area on the right side of the subgraphs where the specialization strategy is dominant increases.
When $\kappa_1=20, 50$ and 100, the area in the subgraphs where $\alpha<1$ is basically dominated by the ordinary strategy. 
In subgraphs with $\kappa_1=1$ and $\kappa_1=10$, we find that the group will choose the ordinary strategy at the final state in region with both $\alpha$ and $M$ being small, and will choose the specialization strategy in other parameter region. 
There are two boundaries between the specialization strategy area and the ordinary strategy area in subgraphs with $\kappa_1=1$ and $\kappa_1=10$, and the position of the two boundaries will lift as $\kappa_1$ increases.
In the subgraphs with two boundaries, when $M$ and $\alpha$ are small or large at the same time, the group tends to adopt the ordinary strategy, while in the other parameter region the group tends to adopt the specialization strategy. 
Fig.~\ref{fig_m_a_e1} shows the simulation results obtained by taking $E=1$ and the other conditions being invariant, which is only slightly different from Fig.~\ref{fig_m_a}. 

The above simulation results in Fig.~\ref{fig_m_a} and Fig.~\ref{fig_m_a_e1} show that, when the allocation parameter is very small (as the situation described in the subfigure with $\kappa_1=0.1$), the specialization strategy is the better choice for competitors.
As the allocation parameter becomes relatively large (as the cases described in the subfigures with $\kappa_1=10$), it is suitable to adopt the ordinary strategy with making less effort when the resource is poor, and as the allocation parameter increases, the range of adopting the ordinary strategy will become larger.
When the allocation parameter is very large (as the cases described in the subfigures with $\kappa_1=20, 50, 100$), choosing the ordinary strategy with less effort is a better choice.

\begin{figure}[!htbp]	
	\centering
	\includegraphics[scale=0.25]{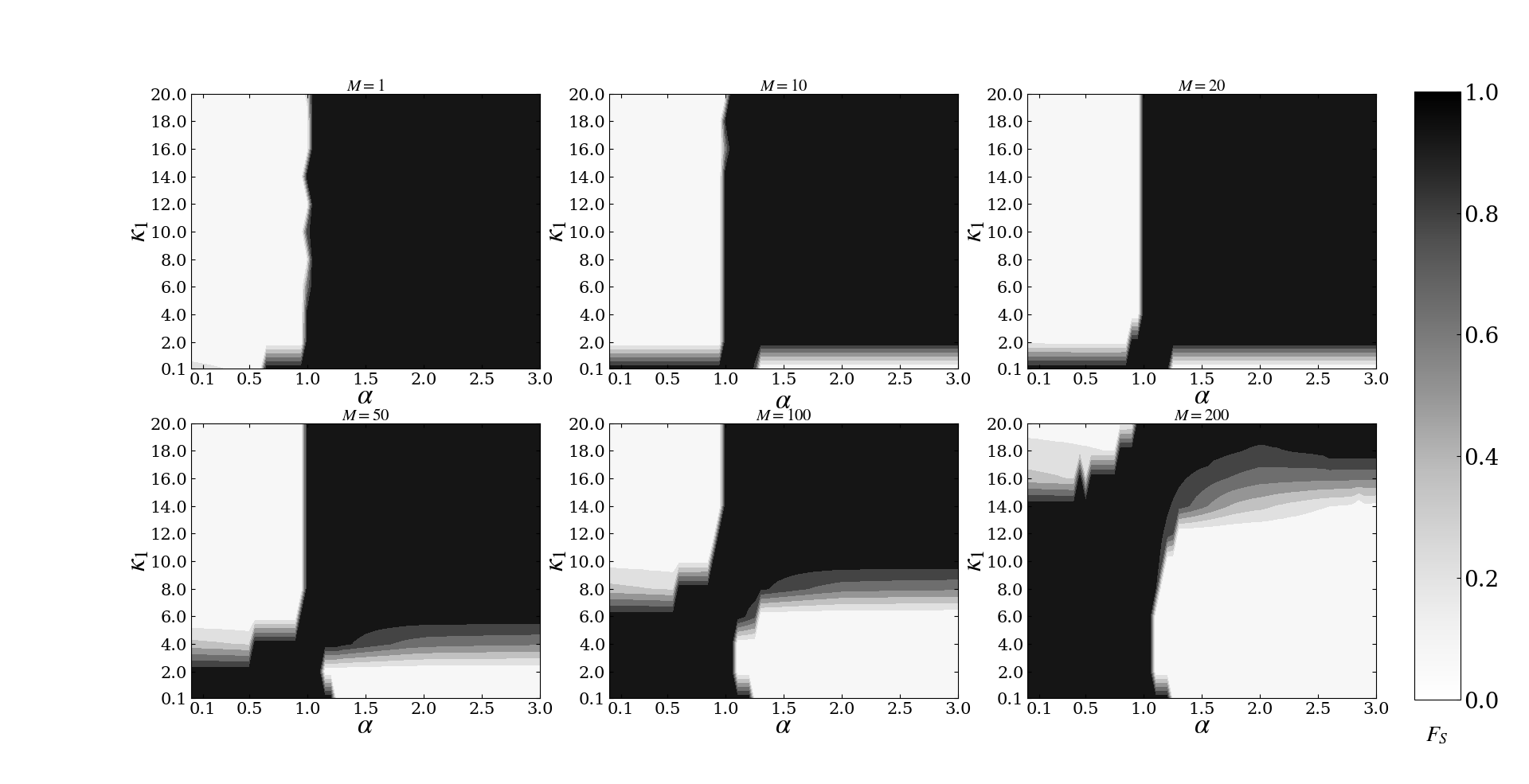}
	\caption{The proportion of the group adopting the specialization strategy, $F_S$, as a bivariate function of the allocation parameter $\kappa_1$ and the ratio $\alpha$ when $E=0.1$ and $M=1, 10, 20, 50, 100, 200$.}
	\label{fig_k_a}
\end{figure}

\begin{figure}[!htbp]	
	\centering
	\includegraphics[scale=0.25]{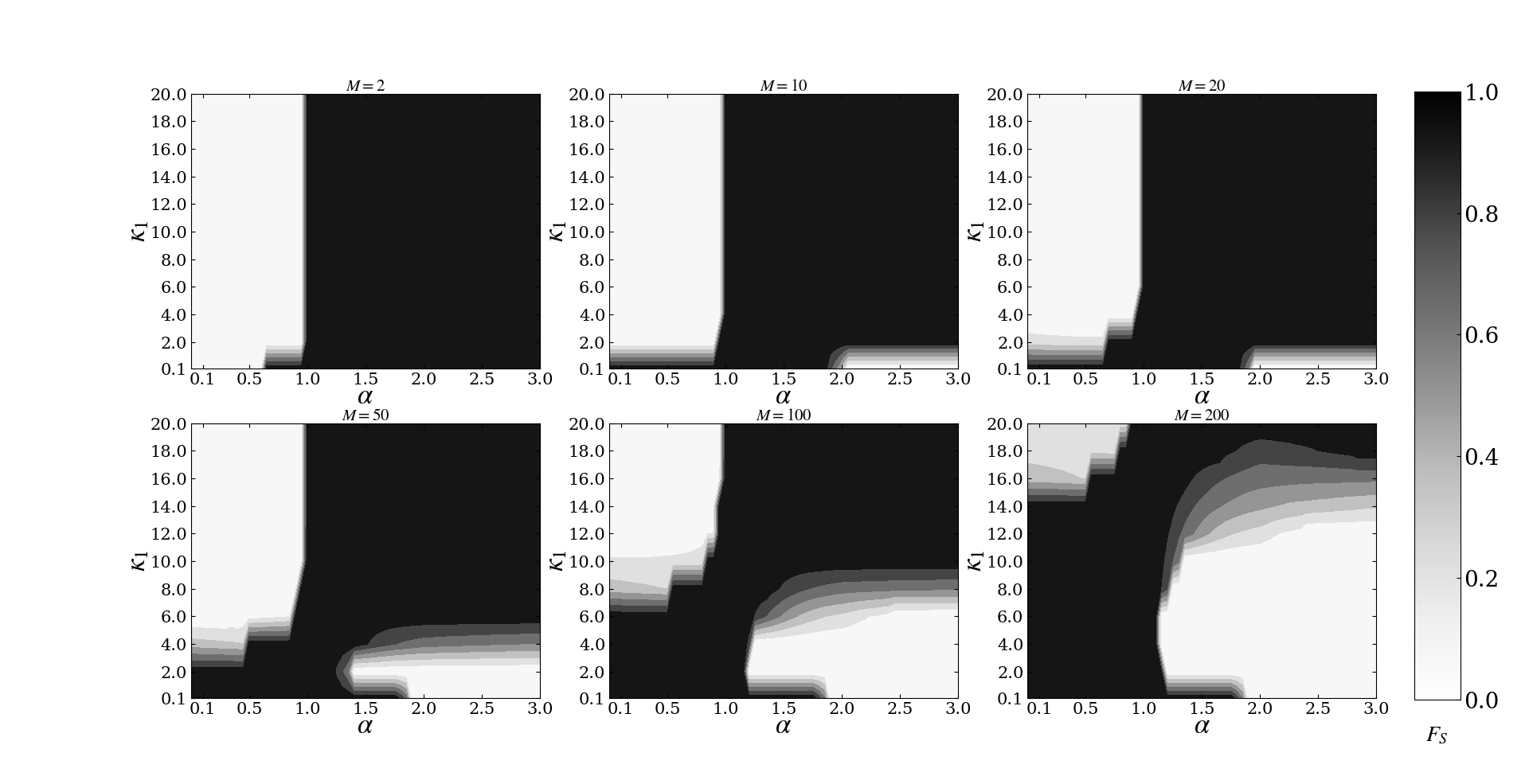}
	\caption{The proportion of the group adopting the specialization strategy, $F_S$, as a bivariate function of the allocation parameter $\kappa_1$ and the ratio $\alpha$ when $E=1$ and $M=1, 10, 20, 50, 100, 200$.}
	\label{fig_k_a_e1}
\end{figure}

In Fig.~\ref{fig_k_a}, we show the relationship of $F_S$ with the allocation parameter $\kappa_1$ and the ratio $\alpha$ when the resourse takes different values and $E=0.1$. 
As shown in the figure, with $M$ increasing, the area on the left side of the subgraphs where the specialization strategy is dominant increases, and the area on the right side of the subgraphs where the specialization strategy is dominant decreases.
In the subgraphs with $M=10, 20, 50, 100, 200$, we find that the group will choose the specialization strategy at the final state in region with both $\alpha$ and $\kappa_1$ being large, and will choose the ordinary strategy in other parameter region.
While in the subgraph with $M=1$, the area where $\alpha<1$ is basically dominated by the ordinary strategy and the other area is dominated by the specialization strategy.  
There are two boundaries between the specialization strategy area and the ordinary strategy area in the subgraphs with $M=10, 20, 50, 100, 200$, and the position of the two boundaries will lift as $M$ increases.
In the subgraphs with two boundaries, when $M$ and $\alpha$ are small or large at the same time, the group tends to adopt the specialization strategy, while the other parameter area tends to adopt the ordinary strategy. 
In order to study the impact of the value of $E$, we also studied the case where $E$ takes different value.
It should be noted that the value of $M$ is set to be greater than $E$.
Fig.~\ref{fig_k_a_e1} shows the simulation results obtained by taking $E=1$ and the other conditions being invariant, which is only slightly different from Fig.~\ref{fig_k_a}.  

The simulations described in Fig.~\ref{fig_k_a} and Fig.~\ref{fig_k_a_e1} illustrate that, when the resource is poor (as the situation described in the subfigure with $M=1$), the ordinary strategy with less effort is a better choice for competitors.
But as the resources become abundant, it is suitable to adopt the specialization strategy when the allocation parameter is very small, and as the resources increase, the range of adopting the specialization will become larger (as the cases described in the subgraphs with $M=10, 20, 50, 100, 200$).

\section{Theoretical analysis}\label{sec_ana}
To further understand the underlying mechanism of the above involution game with specialization strategy, and to avoid illusive results caused by the model setting, we conduct a theoretical analysis on the model as follows.
For agents in the model, the key factor in deciding whether to change their strategies is the difference in payoff resulting from choosing different strategies.
Therefore, we can provide explanations of the simulation results presented in the previous section by analyzing the difference between the payoff of an agent with the specialization strategy and the ordinary strategy in a single game, and the reason for analyzing a single game is that the evolution of the above model is constituted of a large number of games that are essentially not different.

In the analyzed single game, we denote the number of agents adopting the ordinary strategy as $n_0$, the number of agents adopting the specialization strategy $(E, 0)$ as $n_{s_1}$ and the number of agents adopting the specialization strategy $(0, E)$ as $n_{s_2}$, and the three numbers satisfy $n_0+n_{s_1}+n_{s_2}=D+1$.
Since the selection of strategy $(E, 0)$ and $(0, E)$ when an agent decide to take the specialization strategy in the simulation is equal probability, the expected value of $n_{s_1}$ is equal to the expected value of $n_{s_2}$, i.e $\mathbb{E}(n_{s_1})=\mathbb{E}(n_{s_2})$.
In the analysis below, we will take $n_{s_1} \approx n_{s_2}$.

\subsection{All agents making equal effort}
For the scenario of all agents making equal effort, the payoff difference $\Delta \pi$ between the specialization strategy $\pi_S$ and the ordinary strategy $\pi_O$ in a single game is given by
\begin{eqnarray}
	\Delta \pi 
	&=& \pi_S - \pi_O \nonumber
	\\
	&=& \left[\frac{M}{2}\left( \frac{e^{E/\kappa_1}}{n_{s_1}e^{E/\kappa_1}+n_0e^{E/(2\kappa_1)}} + \frac{1}{n_{s_2}e^{E/\kappa_1}+n_0e^{E/(2\kappa_1)}}  \right) - E\right] \nonumber
	\\ 
	& &
	-\left [\frac{M}{2}\left( \frac{e^{E/(2\kappa_1)}}{n_{s_1}e^{E/\kappa_1}+n_0e^{E/(2\kappa_1)}} + \frac{e^{E/(2\kappa_1)}}{n_{s_2}e^{E/\kappa_1}+n_0e^{E/(2\kappa_1)}}  \right) - E \right] \nonumber
	\\
	&=& \frac{M}{2}(x-1)\left[\frac{1}{n_{s_1}x+n_0} - \frac{1}{x(n_{s_2}x+n_0)}\right] \nonumber
	\\
	&\approx& \frac{M}{2}\frac{(x-1)^2}{x(n_{s_1}x+n_0)}
\end{eqnarray}
where $x=e^{E/(2\kappa_1)}$ and $x>1$ as $E/(2\kappa_1)>0$. 
In the approximation of the last step of the above equation, we have taken $n_{s_1}\approx n_{s_2}$.
Though we have assumed that the agent adopts the specialization strategy $(E, 0)$, the approximate result obtained in this case is consistent with the approximate result if the agent takes the specialization strategy $(0, E)$.

As can be seen from the above formula, $\Delta \pi$ approaches 0 when $E/(2\kappa_1)$ is close to 0. 
According to Eq.~\ref{eq_adj}, the probability of agent $i$ adopting agent $j$'s strategy in the next time step is about $1/2$. 
In this case, the value of $F_S$ exhibits randomness, as the green line shown shown in Fig.~\ref{fig_phen}.
When $E/(2\kappa_1)$ is significantly greater than 0, $\Delta \pi$ is significantly greater than 0 too.
Also according to Eq.~\ref{eq_adj}, when agent $i$ adopts the specialization strategy and agent $j$ adopts the ordinary strategy, the probability of agent $i$ transitioning to the ordinary strategy is less than $1/2$; 
when agent $i$'s strategy is the ordinary strategy and agent $j$'s strategy is the specialization strategy, the probability of agent $i$ transforming into the specialization strategy is greater than $1/2$. 
Reflected in the overall evolution of the model, $F_S$ has a tendency to evolve into 1, as the black, orange, blue, brown and red lines shown in Fig.~\ref{fig_phen}.

\subsection{Agents with different strategy making unequal effort}
\begin{figure}[!htbp]
	\centering
	\includegraphics[width=120mm]{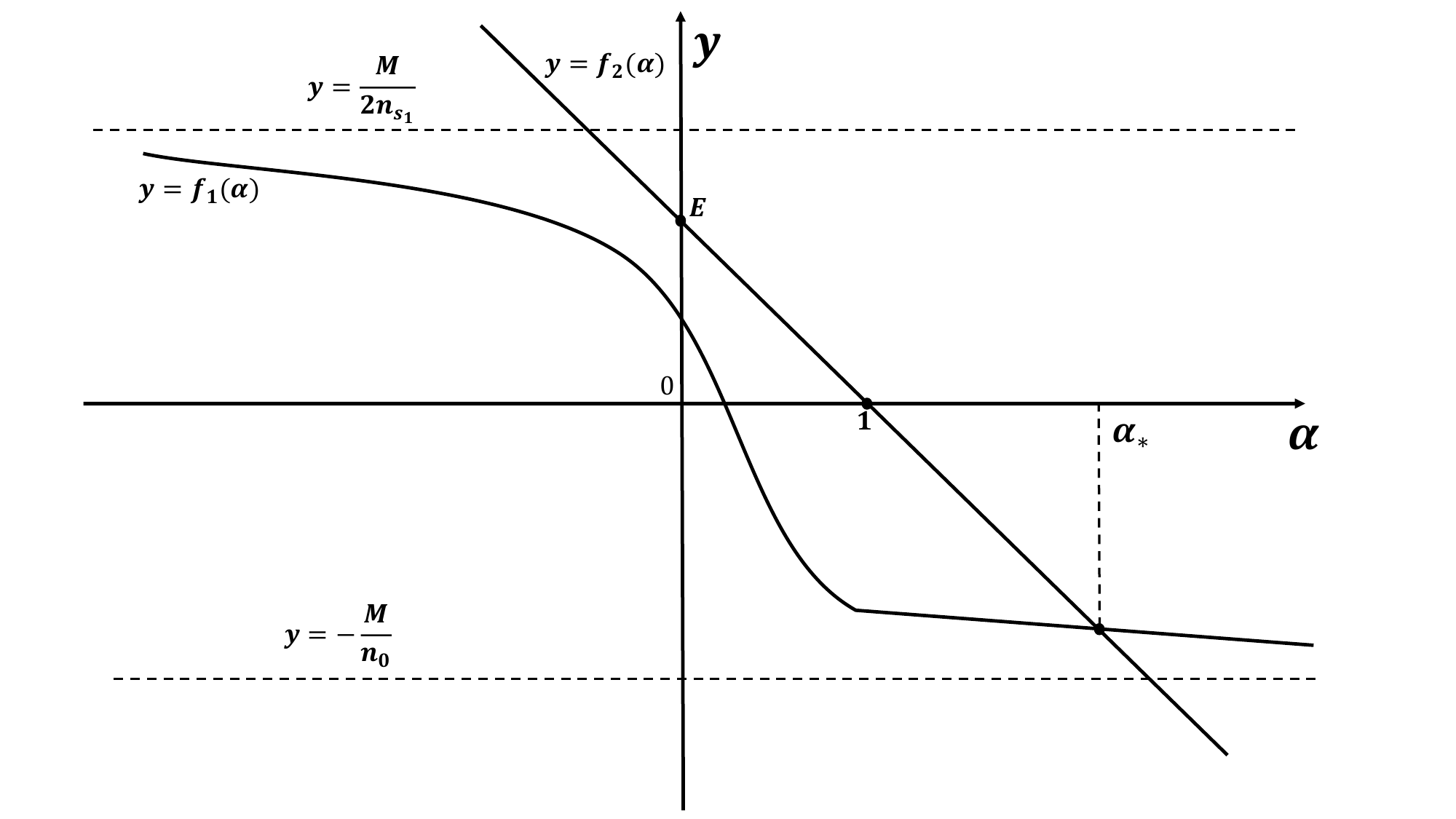}
	\caption{\label{fig_cross1}The sketch map of $y=f_1(\alpha)$ and $y=f_2(\alpha)$ existing only one intersection. The abscissa of the intersection of $y=f_1(\alpha)$ and $y=f_1(\alpha)$ is $\alpha_{*}$, $f_1(\alpha)<f_2(\alpha)$ if $0<\alpha<\alpha_*$, and $f_1(\alpha)>f_2(\alpha)$ if $\alpha>\alpha_*^{'}$.}
\end{figure}

\begin{figure}[!htbp]
	\centering
	\includegraphics[width=120mm]{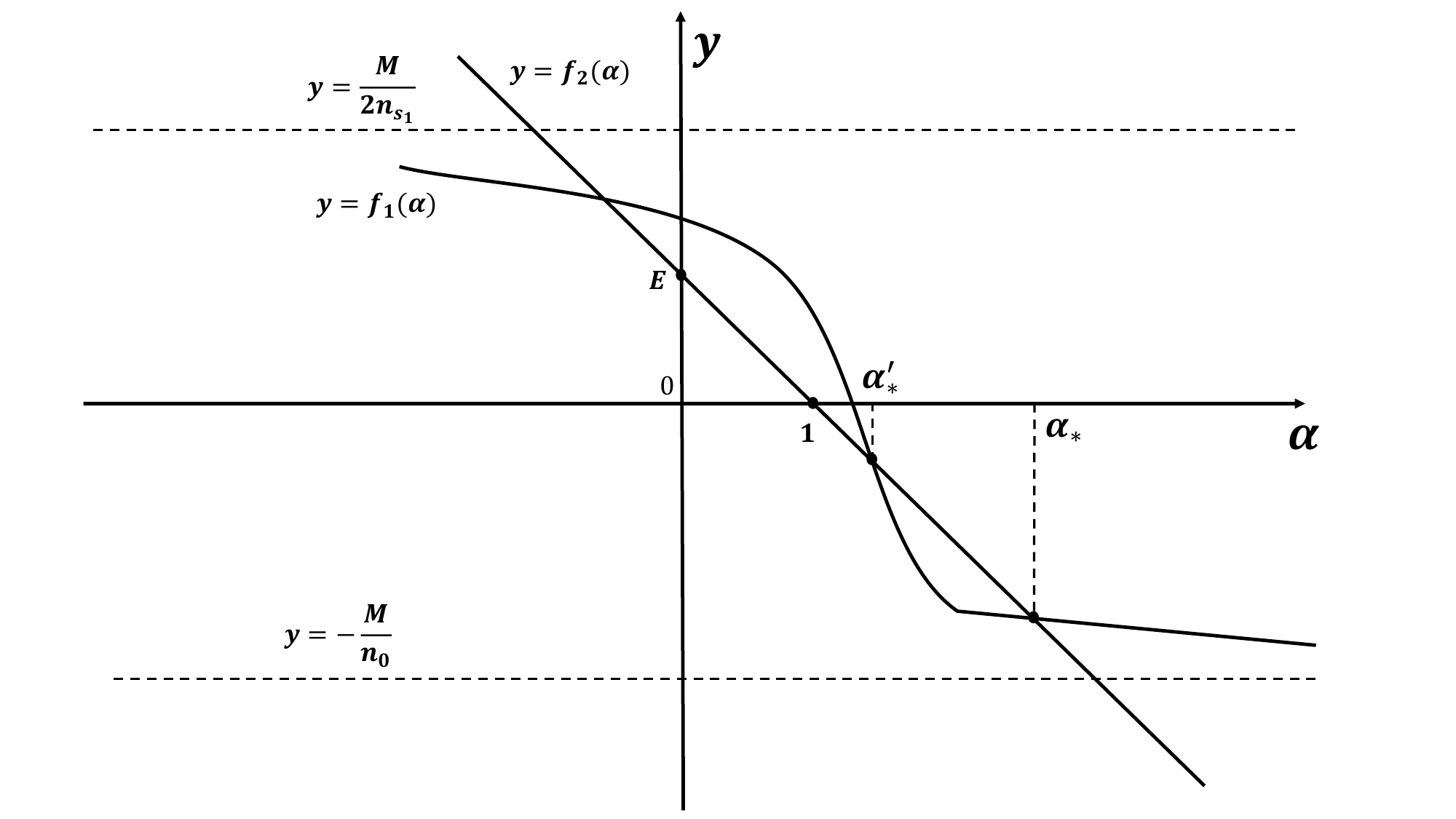}
	\caption{\label{fig_cross2}The sketch map of $y=f_1(\alpha)$ and $y=f_2(\alpha)$ existing two intersections. The abscissas of the intersection of $y=f_1(\alpha)$ and $y=f_1(\alpha)$ are $\alpha_{*}$ and $\alpha_*^{'}$, $f_1(\alpha)>f_2(\alpha)$ if $0<\alpha<\alpha_*^{'}$ or $\alpha>\alpha_*^{'}$, and $f_1(\alpha)<f_2(\alpha)$ if $\alpha_*^{'}<\alpha<\alpha_*$.}
\end{figure}

For the scenario of different strategies with different total effort, the payoff difference $\Delta \pi$ between the specialization strategy $\pi_S$ and the ordinary strategy $\pi_O$ in a single game is given by
\begin{eqnarray}
	\Delta \pi 
	&=& \pi_S - \pi_O \nonumber
	\\
	&=& \left[\frac{M}{2}\left( \frac{e^{E/\kappa_1}}{n_{s_1}e^{E/\kappa_1}+n_0e^{\alpha E/2\kappa_1}} + \frac{1}{n_{s_2}e^{E/\kappa_1}+n_0e^{\alpha E/2\kappa_1}} \right) - E\right] \nonumber
	\\ 
	& &
	-\left[\frac{M}{2}\left( \frac{e^{\alpha E/2\kappa_1}}{n_{s_1}e^{E/\kappa_1}+n_0e^{\alpha E/2\kappa_1}} +
	\frac{e^{\alpha E/2\kappa_1}}{n_{s_2}e^{E/\kappa_1}+n_0e^{\alpha E/2\kappa_1}}\right) - \alpha E\right] \nonumber
	\\	
	&\approx& \frac{M}{2}\left( \frac{x^2+1-2x^{\alpha}}{n_{s_1}x^2+n_0x^{\alpha}}	\right)-(1-\alpha)E \nonumber
	\\
	&=& f_1(\alpha) - f_2(\alpha)
\end{eqnarray}
where $x = e^{E/(2\kappa_1)}>1$, the function $f_1(\alpha)=\frac{M}{2}\left( \frac{x^2+1-2x^{\alpha}}{n_{s_1}x^2+n_0x^{\alpha}}	\right)$ and the function $f_2(\alpha)=(1-\alpha)E$ (in the simulations $\alpha>0$). 
$f_1(\alpha)$ is a monotonically decreasing function and has lower bound $\mathop{f_1(\alpha)}|_{\alpha \to +\infty}=-\frac{2}{n_0}$ and upper bound $\mathop{f_1(\alpha)}|_{\alpha=0}=\frac{x^2-1}{n_{s_1}x^2+n_0}$.
$f_2(\alpha)$ is also a monotonically decreasing function and it is actually a straight line.
The two functions must have a point of intersection, and at most three points of intersection. 
The number of intersection points is determined by the values of $E$, $\kappa_1$ and $M$.
The schematic diagrams of functions $f_1(\alpha)$ and $f_2(\alpha)$ are shown in Fig.~\ref{fig_cross1} and Fig.~\ref{fig_cross2}, which correspond to the intersection of functions $f_1(\alpha)$ and $f_2(\alpha)$ at one point and two points, respectively ($\alpha>0$).

In Fig.~\ref{fig_cross1}, when $\alpha$ is greater than the value of $\alpha_{*}$, $\Delta \pi$ increases with the increase of $\alpha$, that is to say, the advantage of the specialization strategy is expanding. 
Conversely, when $\alpha<\alpha_{*}$, the smaller $\alpha$, the greater the advantage of the ordinary strategy.
In Fig.~\ref{fig_cross2}, we can see that $\Delta\pi >0$ if $0<\alpha<\alpha_*^{'}$,
$\Delta \pi < 0$ if $\alpha_*^{'}<\alpha<\alpha_*$,
and $\Delta\pi >0$ if $\alpha>\alpha_*^{'}$. 

Similar to the analysis in the previous subsection, if $\alpha$ is near the intersection and $\Delta \pi$ is close to zero, 
$F_S$ will reach an non-equilibrium steady state between $0$ and $1$, similar to the simulation results in Reference \cite{wang2022modeling}. 
In this case no randomness presents, and the reason for achieving an non-equilibrium steady state is that the ordinary strategy and the specialization strategy are no longer similar when $\alpha$ is significantly greater than 0,
which is different from the case of all agents making equal effort.
If $\Delta \pi$ is significantly bigger than 0,  $F_S$ will stabilize at 0 or 1, which is the same as the case of all agents making equal effort.

Different from the case of all agents making equal effort, $\Delta \pi$ may exist two or three zero points.
If there is only one zero point in the parameter range, a boundary will appear in the simulation results of $F_S$, corresponding to the simulation results in subgraphs with $\kappa_1\geq 20$ in Figs.~\ref{fig_m_a}-\ref{fig_m_a_e1}, and subgraphs with $M=1$ and $M=2$ in Figs.~\ref{fig_k_a}-\ref{fig_k_a_e1}. 
If there are two zero points in the parameter range, then two dividing lines will appear in the simulation results regarding of $F_S$, corresponding to the results in subgraphs with $\kappa_1\leq 10$ in Fig.~\ref{fig_m_a}-\ref{fig_m_a_e1}, and subgraphs with $M\geq 10$ in Figs.~\ref{fig_k_a}-\ref{fig_k_a_e1}. 

The above analysis is consistent with the simulation results, and can mine the underlying mechanism of the simulation results.

\section{Discussion and Conclusion}\label{sec_con}
This study constructs the involution game with specialization strategy by redefining the existing model, and considers two situations in which the effort of the specialization strategy and the ordinary strategy is equal and unequal respectively. 
The main conclusion is as below.

When the effort of the two strategies is equal, and the ratio of the allocation parameter to the effort ($E/\kappa_1$) exceeds a certain degree, the group will generally adopt the specialization strategy, and the lower the ratio, the richer the resource is, the more inclined the group to adopt the specialization strategy. 
when $E/\kappa_1$ is approaching 0, the proportion of the specialization strategy adoption will change dramatically with the resource.
In the real world, the difference in effort between individuals is not significant in most cases, so adopting a specialization strategy is better under the same level of effort.

When the effort of the two strategies is different and the allocation parameter is very small, the specialization strategy is the better choice for competitors. As the allocation parameter becomes relatively large, it is suitable to adopt the ordinary strategy with making less effort when the resource is poor, and as the allocation parameter increases, the parameter range of adopting the ordinary strategy will become larger. When the allocation parameter is very large, choosing the ordinary strategy with less effort becomes a better choice.

When the effort of the two strategies is different and the resource is poor, the ordinary strategy with less effort is a better choice for competitors.
But as the resources become abundant, it is suitable to adopt the specialization strategy when the allocation parameter is very small, and as the resources increase, the parameter range of adopting the specialization will become larger.
The above conclusions have some implications for how to make choices about strategies in different competitive environments, 
according to them we can make a choice between the specialization strategy and the ordinary strategy basing on the specific competitive situations.

Our model can also explain why the division of labor in human society is becoming increasingly sophisticated from the perspective of strategy selection.
In the process of human social development, human beings have always faced the problems of resource scarcity (low resource $M$) and fierce competition (low allocation parameter $\kappa_1$).
This situation can correspond to some subfigures in Fig.~\ref{fig_m_a}, Fig.~\ref{fig_m_a_e1}, Fig.~\ref{fig_k_a} and Fig.~\ref{fig_k_a_e1}, we can see that in relevant subgraphs the group tends to adopt the specialization strategy.
When some certain members or units of society gain advantages in competitions by adopting the specialization strategy, other members or units of society will adopt the specialization strategy with some possibilities in the upcoming competitions. After multiple rounds of competitions, the specialization strategy will become increasingly common in human society.
In the real world, we can easily find counterparts. For example, for various fast food restaurants in a region, comprehensive restaurants are becoming increasingly rare, most restaurants owners adopt the specialization strategy and focus their limited resources on several kinds of food.
Especially on Chinese food delivery platforms, almost all restaurants only provide specific types of food.
Another general example is that, the refinement of the division of labor in various manufacturing industries can also be illustrated by our proposed model.

But it must be pointed out that, the reasons for ``division of labor" may have different interpretations in different specific questions and contexts.
Our work is a good attempt to explain this phenomenon using evolutionary game models, but to fully explain it from the evolutionary game perspective, more in-depth research is needed, such as how to determine the model's parameters from real social scenarios.

\section*{Acknowledgments}
\begin{sloppypar}
This work is supported by Wuhan East Lake High-Tech Development Zone (also known as the Optics Valley of China, or OVC) National Comprehensive Experimental Base for Governance of Intelligent Society and the National Natural Science Foundation of China (Grant No. 71932008).
\end{sloppypar}

\bibliographystyle{ieeetr}
\bibliography{reference}
\end{document}